\documentclass[3p,review,pdftex,authoryear]{elsarticle}

\input{pkg}
\newcommand{\pD}[1]{\ensuremath{\partial_{#1}}\xspace}

\newcommand{\Ra}{\ensuremath{\mathrm{Ra}}\xspace}
\newcommand{\Pra}{\ensuremath{\mathrm{Pr}}\xspace}

\newcommand{\wi}{\ensuremath{\tilde{w}_1}\xspace}

\newcommand{\Td}{\ensuremath{\theta^{\ast}}\xspace}
\newcommand{\Tmax}{\ensuremath{\theta^{\ast}_\text{max}}\xspace}
\newcommand{\Tmin}{\ensuremath{\theta^{\ast}_\text{min}}\xspace}
\newcommand{\Tdo}{\ensuremath{\theta_0^{\ast}}\xspace}

\newcommand{\uavec}{\ensuremath{\mathbf{\tilde{u}}}\xspace}
\newcommand{\uaveco}{\ensuremath{\mathbf{\tilde{u}}_0}\xspace}
\newcommand{\uaveci}{\ensuremath{\mathbf{\tilde{u}}_1}\xspace}

\newcommand{\wdi}{\ensuremath{w_1^{\ast}}\xspace}
\newcommand{\td}{\ensuremath{t^{\ast}}\xspace}

\newcommand{\zd}{\ensuremath{z^{\ast}}\xspace}

\newcommand{\Tao}{\ensuremath{\tilde{\theta}_0}\xspace}
\newcommand{\Tai}{\ensuremath{\tilde{\theta}_1}\xspace}
\newcommand{\uao}{\ensuremath{\tilde{u}_0}\xspace}
\newcommand{\vao}{\ensuremath{\tilde{v}_0}\xspace}
\newcommand{\wao}{\ensuremath{\tilde{w}_0}\xspace}
\newcommand{\uai}{\ensuremath{\tilde{u}_1}\xspace}
\newcommand{\vai}{\ensuremath{\tilde{v}_1}\xspace}
\newcommand{\wai}{\ensuremath{\tilde{w}_1}\xspace}
\newcommand{\ta}{\ensuremath{t}\xspace}
\newcommand{\xa}{\ensuremath{x}\xspace}
\newcommand{\ya}{\ensuremath{y}\xspace}
\newcommand{\za}{\ensuremath{z}\xspace}
\newcommand{\kx}{\ensuremath{a_x}}
\newcommand{\ky}{\ensuremath{a_y}}
\newcommand{\ka}{\ensuremath{a}}
\newcommand{\kac}{\ensuremath{a_c}}
\newcommand{\laplacea}{\ensuremath{\tilde{\Delta}}\xspace}
\newcommand{\Tso}{\ensuremath{\theta_0}\xspace}
\newcommand{\Tsi}{\ensuremath{\theta_1}\xspace}

\newcommand{\wsi}{\ensuremath{w_1}\xspace}
\newcommand{\zs}{\ensuremath{\zeta}\xspace}
\newcommand{\ts}{\ensuremath{\tau}\xspace}

\newcommand{\TSI}{\ensuremath{\textsc{dp}}\xspace}
\newcommand{\Ras}{\ensuremath{\mathrm{Ra}_{\ts}}\xspace}
\newcommand{\ks}{\ensuremath{a_{\ts}}}

\newcommand{\thos}{\ensuremath{\theta_0}\xspace}
\newcommand{\tas}{\ensuremath{\tau}\xspace}
\newcommand{\tc}{\ensuremath{\tas_c}\xspace}
\newcommand{\tm}{\ensuremath{\tas_m}\xspace}
\newcommand{\tu}{\ensuremath{\tas_u}\xspace}

\newcommand{\dd}{\mathrm{d}\xspace}
\newcommand{\DD}{\mathrm{D}\xspace}

\newcommand{\zas}{\ensuremath{\zeta}\xspace}

\newcommand{\Ta}{\ensuremath{\tilde{\theta}}\xspace}

\newcommand{\erfc}{\operatorname{erfc}}
\newcommand{\erf}{\operatorname{erf}}
\newcommand{\ii}{\operatorname{i}}

\newcommand{\gvec}{\ensuremath{\mathbf{g}}}

\journal{Ihle, C. F., \& Niño, Y. (2006). The onset of nonpenetrative convection in a suddenly cooled layer of fluid. International Journal of Heat and Mass Transfer, 49(7-8), 1442-1451.\url{https://doi.org/10.1016/j.ijheatmasstransfer.2005.09.025}}

\begin{document}

\begin{frontmatter}
\title{The onset of nonpenetrative convection in a suddenly cooled layer of fluid}
\author{Christian F. Ihle}
\address{Department of Mining Engineering, Universidad de Chile}\ead{cfihle@uchile.cl}%

\author{Yarko Ni\~{n}o}
\address{Department of Civil Engineering,
Universidad de Chile}%
\date{2 September 2005}
\begin{abstract}
Conditions for the onset of nonpenetrative convection in a
horizontal Boussinesq fluid layer subject to a step change in
temperature are studied using propagation theory. A wide range 
of Prandtl numbers and two different kinematic boundary conditions 
are considered. It is shown that for high Rayleigh numbers, critical 
conditions for the onset of convective motion reproduce exactly those 
for the unsteady Rayleigh-B\'enard instability. Present results extend 
those of previous research
and show a tendency of the rigid-rigid and free-rigid 
critical curves to converge for low Prandtl numbers. Comparison between 
present and previously reported results on critical conditions for the onset
of instabilities and onset time using different methods yields
good agreement on a middle to high Prandtl number range. A ratio of $10$ 
between experimentally measured and theoretically predicted onset times is 
suggested for stress-free bounded systems.
\end{abstract}
\begin{keyword}
Buoyancy-driven instability\sep Critical time\sep Nonpenetrative 
convection\sep Prandtl number\sep
Propagation theory\sep Rayleigh number\sep Isothermal heating.%
\end{keyword}
\end{frontmatter}

\section*{Nomenclature}
\begin{tabbing}
\makebox[10ex][l]{$a_s$, $b_s$}	\= coefficients of Eqs.~\eqref{e:corr_dp_Ra} 
		and~\eqref{e:corr_dp_a}, respectively,
		$s=1,\ldots,5$	(integer), or $\infty$\\
$(\kx,\ky)$ \> dimensionless horizontal wavevector \\
$\ka$ \> dimensionless horizontal wavenumber, $\sqrt{\kx^2+\ky^2}$\\
$C$         \> concentration [kmol$/$m$^3$] or designation of constant value\\
$C_p$       \> specific heat of the fluid at constant pressure [J\,kg$^{-1}$\,K$^{-1}$]\\
$D$		\> mass diffusion coefficient [m$^2$/s]\\
$\DD(\cdot)$\> ordinary derivative with respect to $\zs$, $\mathrm{d}(\cdot)/\mathrm{d}\zs$\\
$\pD{\chi(\psi)}(\cdot)$	\> partial derivative, $\partial(\cdot)/\partial\chi$ or 
				$\partial^2(\cdot)/\partial\chi\psi$\\
$\TSI$\> 		deep pool acronym\\
$\gvec$     \> gravity vector (pointing in the direction of $\za$
			axis) [m\,s$^{-2}$]\\
$k$         \> thermal conductivity of the fluid [W\,m$^{-1}$K$^{-1}$]\\
$L$         \> depth of the fluid layer [m]\\
$\Pra$      \> Prandtl number, $\nu\alpha^{-1}$\\
$r$ \> slope of the geometrical sequence to extrapolate critical $\Ras$ values\\
$\ta$       \> time (dimensionless if no superscript)\\
$\mathbf{u}$	\> velocity, $(u,v,w)$  (dimensional or not depending on the superscript.\\
		\> $w$ is $\zs$-dependent if no superscript is present)\\
$\Ra$ \> Rayleigh number, $g\beta(\Tmax-\Tmin)L^3\nu^{-1}\alpha^{-1}$\\
$\Ras$  \> $\tau$-dependent Rayleigh number, $\ts^{3/2}\Ra$\\
TBL \> thermal boundary layer acronym\\
$(x,y,z)$	\> Cartesian coordinates (dimensionless if no superscript is present)\\
\end{tabbing}
\nopagebreak

\subsection*{Greek letters}
\begin{tabbing}
\makebox[5ex][l]{$\alpha$}  \= thermal diffusivity of the fluid [m$^2$\,s$^{-1}$]\\
$\beta$     \> thermal expansion coefficient [K$^{-1}$]\\
$\lambda$ \> relative difference coefficient, $\max_{\zs}\{100\times|1-\Tso/\Tso{}_{\TSI}|\}$\\
$\delta_{\theta}$  \> dimensionless thermal penetration depth\\
$\Delta$	\> laplacian operator (dimensional or not depending
		on the superscript)\\
$\Delta_1$	\> horizontal laplacian operator (dimensional or not depending
		on the superscript)\\
$\gamma$	\> concentration coefficient of expansion [kg$/$kmol]\\
$\zs$		\> self-similar vertical coordinate, $\za/\sqrt{\ta}$\\
$\theta$	\> temperature (self-similar if no superscript, 
			otherwise dimensional or dimensionless)\\
$\nu$       \> kinematic viscosity of the fluid [m$^2$\,s$^{-1}$]\\
$\sigma$    \> temporal growth rate for disturbances\\
$\ts$ \> definition for time in the self-similar framework, $\ts=\ta$\\
\end{tabbing}

\subsection*{Subscripts}
\begin{tabbing}
\makebox[8ex][l]{$0$} \= base state\\
$1$ \> disturbance, or correlative assignment to constant\\
$2$--$5$	\> correlative assignment to constant\\
$b$		\> bulk\\
$c$		\> critical state\\
\TSI		\> deep pool assumption: $\zs$-only dependence\\
linear		\> linear boundary forcing\\
$\infty$	\> infinite Prandtl number\\
$m$		\> experimental detection\\
min		\> minimal condition\\
max		\> maximal condition\\
$u$		\> thermal advection dominance over pure diffusivity\\
step		\> step boundary forcing\\
$\tau$		\> $\tau$-dependent variable: $\Gamma_{\tau}\equiv\tau^{\phi}\Gamma$
\end{tabbing}

\subsection*{Superscripts}
\begin{tabbing}
\makebox[8ex][l]{$\ast$}  \= dimensional length, temperature,
time, velocity or
differential operator\\
$\sim$  \> dimensionless temperature, velocity or differential operator\\
--      \> root-mean-square\\
\end{tabbing}

\section{Introduction}

Nonpenetrative convection is defined by~\citet{Adrian86ef} as 
the unstable flow field that derives from the existence of a fluid
layer heated from below (or cooled from above) with adiabatic
top (or bottom if cooled from above), resembling 
B\'enard convection~\citep{Rayleigh16pm}. This thermal boundary 
condition precludes the existence of a steady flow regime.
Nonpenetrative convection represents a reasonable assumption 
in a variety of physical problems, that range from ventilation 
and air conditioning (like, for instance, cold storage rooms and
warehouses with poor insulation from one side) to earth sciences,
particularly regarding the dynamics of the planetary boundary
layer~\citep{Stull88book}. In this paper, the attention is focused
on the study of conditions for the onset of impulsively generated
nonpenetrative convection. Here, the base state of the system to be 
perturbed, at 
difference from the one that gives rise to B\'enard 
convection~\citep{Drazin81book}, is unsteady
due to the existence of a thermally diffusive state whose temporal
rate of change is high at the very beginning of the evolution. 
Hence, a stability model able to deal with this difficulty is to 
be considered.

After early approaches to the analysis of the stability of unsteady systems 
(e.g. `frozen time' and `quasi-static' models,
reviewed by~\citet{Gresho71ijhmt} and~\citet{Homsy73jfm}, respectively), 
the study of the onset of manifest convection in 
high Rayleigh number fluid layers impulsively heated 
or cooled began with~\citet{Foster65b_pf}, who used an 
initial value technique, so-called `amplification model',
which considers a transient evolution of the base state. 
In this case, disturbances that cause the onset of convection are 
assumed to occur only initially. The major
drawback of this method is that determination of amplification
requires the knowledge of amplitudes of initial disturbances, for
all the wavelengths present on the eigenfunction expansion. As
this is impossible, Foster's approach consisted of a heuristic
procedure that combined the assumption of several disturbance
patterns along with experimental observations~\citep{Foster69pf}.
Using a different approach,~\citet{Jhaveri82jfm} and~\citet{Kim86ijhmt} 
used random forcing functions to solve an initial value problem to find 
the onset times, both for step and ramp-heated systems of high Rayleigh
numbers, suggesting a definition of the onset time as that corresponding 
to a certain excess of the computed Nusselt number with respect to the 
purely conductive one.

More recently,~\citet{Kim99ijhmt} studied the impulsively
driven Rayleigh-B\'enard problem with initial stratification, using
the method called by these authors `propagation theory'
\citep{Choi88proc,Kim96kjce,Kang97pf}.
Its basis lies on the assumption that most of the disturbances are
confined within the thermal penetration depth, which is considered
as a length scale, leading to the transformation of the linearized
equations into self-similar forms. 
In more recent
contributions,~\citet{Chung04kjce} and~\citet{Choi04ijts} suggest
new definitions for onset times, taking into account nonlinear
effects that come from the numerical simulation of the unsteady
Rayleigh-B\'enard problem and compare them with results obtained using 
propagation theory. In the latter work, the influence of 
initial stratification on the distribution of the mentioned time 
scales is analyzed.

In an experimental context,~\citet{Spangenberg61pf}
studied the onset of evaporative convection
using Schlieren photography techniques, while~\citet{Foster65pf}, 
by means of radiometry, showed that surface 
temperature in suddenly cooled evaporative systems evolves in a linear
fashion.~\citet{Plevan66aj}, \citet{Blair69jfm} and later~\citet{Tan92ces}, 
measured onset times in non-evaporative systems whose stability depends on the 
concentration of gases into water. Their results,
in the context of the present research, are commented in 
Section~\ref{s:results_and_discussion}.
\citet{Goldstein95jhtta},
studied the onset of convection on a thick fluid layer heated
impulsively from below. Their work presents also an extensive
review of literature focused on the transient features of natural
convection.  

In this paper, propagation theory was the chosen stability method
to assess the onset of nonpenetrative convective motion. For high
thermal perturbations, it is shown that this phenomenon behaves
the same as the onset of unsteady Rayleigh-B\'enard convection. This
result allows for a side by side comparison of present computations
with numerical and experimental results reported in the context of
the latter problem.  Some new findings in that regard are presented
and discussed as well.

\section{Problem description}

\begin{figure}[!h]
\begin{center}
\includegraphics[width=0.5\textwidth]{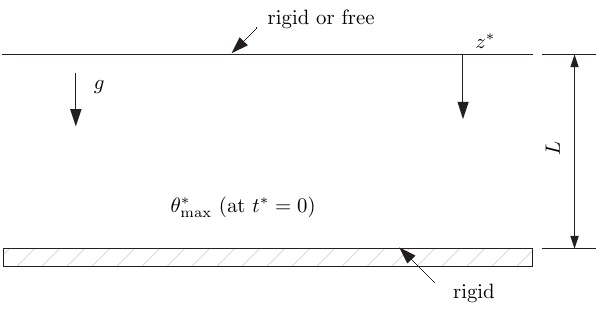}
\end{center}
\caption{Problem configuration. $\zd$ axis points
downward.}\label{f:conf}
\end{figure}

An initially quiescent horizontal fluid layer, well mixed at
temperature $\Td=\Tmax$, infinite on its horizontal dimension but
finite, with height $L$, on the vertical axis $\zd$, is suddenly
cooled, by dropping its surface temperature, at time $\td=0$ and
$\zd=0$, to $\Td=\Tmin$. Surface is to be kept at this lower
temperature for $\td>0$ (Fig.~\ref{f:conf}).

For high enough temperature step: $\delta\Td=\Tmax-\Tmin$, a
buoyancy-driven circulation is induced. This problem can be
modeled using continuity, Navier-Stokes and energy equations on a
Boussinesq fluid, with no heat sources present. Surface tension effects 
in the free-rigid case are neglected in the present study. This assumption 
is reasonable in the present context, as shown experimentally
by~\citet{Davenport74ijhmt} in the case of linearly heated deep
reservoirs.
Scales to be used are $L$ to form dimensionless coordinates
($\xa$, $\ya$, and $\za$), $L^2\alpha^{-1}$ to form dimensionless
time, $\ta$, $\alpha L^{-1}$ to form dimensionless velocity
base state and perturbations, $(\uao,\vao,\wao)$ and 
$(\uai,\vai,\wai)$, respectively. $\nu\alpha g^{-1}\beta^{-1}L^{-3}$,
to form dimensionless temperature 
perturbation, $\Tai$, whereas the dimensionless base temperature, 
$\Tao$, is scaled to range between $0$ and $1$: 
$\Tao=(\Tdo-\Tmin)(\Tmax-\Tmin)^{-1}$. $\alpha$, $\nu$ and 
$\beta$ are the thermal diffusivity, kinematic viscosity, and thermal 
expansion coefficient of the fluid, respectively. $g$ is the magnitude of
the gravity vector, which points in the same direction of the $\za$ axis.
In the latter expressions, the 
subscript $0$ refers to the base state and $1$ to the perturbed one. 

A first order expansion for the dimensionless 
temperature and velocity is considered, with the form $\Ta=\Tao-\Tai$, 
and $\uavec = \uaveco+\uaveci=\uaveci=(\uai,\vai,\wai)$, respectively. The 
minus sign on the expansion for temperature means that positive perturbations 
have a cooling effect. The base state is that of a horizontally infinite,
quiescent fluid layer. Neglecting second order terms, the following set
of equations is obtained for the vertical velocity and temperature perturbations:
\begin{subequations}\label{e:nodim_stab}
\begin{gather}
\left(\frac{1}{\Pra}\pD{\ta}- \laplacea  \right)\laplacea\wai = \laplacea_1\Tai\label{e:nodim_stab1}\\
\pD{\ta}\Tai-\Ra\,\wai\pD{\za}\Tao=\laplacea\Tai\label{e:nodim_stab2},
\end{gather}
\end{subequations}
where $\Pra=\nu\alpha^{-1}$ corresponds to the Prandtl number and
$\Ra=g\beta(\Tmax-\Tmin)L^3\nu^{-1}\alpha^{-1}$ corresponds to a
Rayleigh  number based on the overall temperature step, 
$\laplacea\equiv\pD{\xa\xa}+\pD{\ya\ya}+\pD{\za\za}$ and 
$\laplacea_1\equiv\laplacea-\pD{\za\za}$, 
provided the dimensionless equation for the base state
is satisfied:
\begin{subequations}\label{e:a_base_state}
\begin{gather}
    \pD{\ta}\Tao = \pD{\za\za}\Tao\\
    \Tao(\ta=0,\za)=1,\quad\Tao(\ta>0,\za=0)= \pD{\za}\Tao(\ta\geq0,\za=1)=0.
\end{gather}
\end{subequations}%
The derivation of stability equations using propagation theory is
analogous to that of~\citet{Kang97pf} and~\citet{Yang02pf}. Hence,
only the essential steps are given here. In propagation theory it
is stated that, for the case of thermal convection in systems
where instabilities are confined mainly into the thermal boundary layer 
(TBL), a balance between viscous and buoyant forces can be made, such 
that it is possible to scale dimensionless vertical velocity perturbations
with time as $|\wai\Tai^{-1}|\sim\delta_{\theta}^2$, where
$\delta_{\theta}\propto\sqrt{\ta}$ is the dimensionless thermal
penetration depth. From the latter relation and dimensional
analysis it can also be inferred that~\citep{Yang02pf}:
\begin{equation}\label{e:w_scale}
\left[\Tai\left(\za,\ta\right),\wai\left(\za,\ta\right)\right] =
    \left[\ta^n\Tsi\left(\za\bigl/\sqrt{\ta}\right),\ta^{n+1}\wsi\left(\za\bigl/\sqrt{\ta}\right)\right],
\end{equation}
where $n$ is a parameter. Now, stability equations are represented
in a new coordinate system defined as $\left(\ta,\zs=\za\bigl/\sqrt{\ta}\right)$,
instead of $\left(\ta,\za\right)$, while $\Tai$ and $\wai$ turn to $\Tsi$ and
$\wsi$ in the newly defined system. To avoid confusion, $\ta$ will
be defined as $\ts$. The present criterion for the setting of $n$
is to find the lowest possible onset times from the characteristic
problem. To this purpose, it must be set to zero~\citep{Yang02pf}.
Additionally,~\citet{Choi04ijts} and~\citet{Chung04kjce} argue
that this condition can be also derived from the assumption that
the onset time  occurs when the growth rates of the
root-mean-square values of the base state temperature and of the
temperature perturbations are equal. In the context of a system
with an imposed heat flux, the latter assumption leads to a different $n$
value of $1/2$~\citep{Choi04hmt,Choi04ijhmt}.

Eqs.~\eqref{e:nodim_stab} are cyclic in
the horizontal plane. Then, modes with wavenumbers $\kx$ and $\ky$ for
the $\xa$
and $\ya$ axis, respectively, are considered.
Introducing~\eqref{e:w_scale}$\times\exp[\ii(\kx\xa+\ky\ya)]$ in
the latter system, noting that
$\pD{\ts}(\cdot)=-(2\ts)^{-1}\pD{\zs}(\cdot)$ and that
$\pD{\za}(\cdot)=\ts^{-1/2}\pD{\zs}(\cdot)$, the set of stability
equations to be solved is:
\begin{subequations}\label{e:stab_eqn}
\begin{gather}
\left[\left(\DD^2-\ks^2 \right)^2+\frac{1}{2\Pra}\left(\zas \DD^3-
    \ks^2\zs \DD + 2\ks^2\right)\right]\wsi - \ks^2\Tsi = 0\label{e:stab_eqn1}\\
\left(\DD^2+\frac12\zs \DD - \ks^2 \right)\Tsi+\wsi\Ras
\DD\thos=0,\label{e:stab_eqn2}
\end{gather}
\end{subequations}
where $\DD^n(\cdot)=\dd^n(\cdot)/\dd{\zs^n}$,  $\ks =
\ts^{1/2}\sqrt{\kx^2+\ky^2}$ and $\Ras=\ts^{3/2}\Ra$.

The scaling assumed here, which considers the hypothesis that
disturbances are confined mainly into a thermal penetration depth,
makes Eqs.~\eqref{e:stab_eqn} valid for
small values of time only. In this case, the base state for
temperature, $\Tso$, can be expressed as a function exclusively of
$\zs$. This kind of system, representative of a thermally
semi-infinite one, is commonly named `deep pool' system (the
acronym~\TSI will be adopted hereafter). Its TBL is small compared
with the thickness of the fluid layer. 

For large values of $\ts$, when equations are not self-similar
anymore, it has been shown~\citep{Yang02pf,Kim02tcfd} that
eigenvalues for~\eqref{e:stab_eqn}
can still be found. In those works, 
it was also shown that asymptotic convergence in time to results 
obtained with the frozen time model is achieved. However, the validity at
intermediate values of time of the thermal scaling proposed here
is not clear. 
Regarding this topic, an analysis on the validity of
this model, in the context of nonpenetrative convection, is being
presently prepared~\citep{Ihle05xxx}.
For the system with no-slip top and 
bottom surfaces (named herein as the rigid-rigid case), boundary conditions 
for the perturbed quantities are:
\begin{subequations}\label{e:rr}
\begin{gather}
\Tsi=\wsi=\DD\wsi=0\quad\text{in }\zs=0\label{e:rr1}\\
\DD\Tsi=\wsi=\DD\wsi=0\quad\text{in }\zs=1/\sqrt{\ts}.\label{e:rr2}
\end{gather}
\end{subequations}
In the case with stress-free top and no-slip bottom (defined also as the
free-rigid case), boundary conditions which are to be applied
to Eqs.~\eqref{e:stab_eqn} are:
\begin{subequations}\label{e:fr}
\begin{gather}
\Tsi=\wsi=\DD^2\wsi=0\quad\text{in }\zs=0\label{e:fr1}\\
\DD\Tsi=\wsi=\DD\wsi=0\quad\text{in }\zs=1/\sqrt{\ts}\label{e:fr2}.
\end{gather}
\end{subequations}
The marginal stability problem to be considered is to solve:
$\min_{\ks}\Ras$, where $\ks$ and $\Ras$
satisfy~\eqref{e:stab_eqn}, with boundary
conditions~\eqref{e:rr} or~\eqref{e:fr}, in the rigid-rigid and 
free-rigid cases, respectively. This procedure is to be applied to 
the self-similar system, valid for small values of time. 
Under this condition, the present definition of the Rayleigh number is 
the same as the one used in the classical Rayleigh-B\'enard problem, based 
on the temperature difference between the top and bottom horizontal
boundaries, since here, the bottom boundary holds its higher
temperature throughout the whole lapse of time during which the
present stability model is valid.

\section{Solution method}

Eqs.~\eqref{e:stab_eqn} and the boundary
conditions~\eqref{e:rr} and \eqref{e:fr} are homogeneous. Then, the
value of $\DD^2\wsi(0)$ and $\DD\wsi(0)$ can be assigned
arbitrarily in the rigid-rigid and free-rigid cases,
respectively~\citep{Kang97pf,Kim02tcfd}. To solve the problem
posed in the previous section, a solver based on the shooting
method using a fourth order Runge-Kutta numerical integration
formula was implemented. Convergence to minima was achieved using
a Newton-Raphson scheme. Validation of the numerical
implementation was done by analyzing the classical rigid-rigid
Rayleigh-B\'enard problem with a step change in the bottom
temperature~\citep{Kim99ijhmt}, using Eqs.~\eqref{e:stab_eqn}. 
Monotonic, albeit slow convergence for
increasing time, close to the well known value of the critical
Rayleigh number of $1708$ was found for different Prandtl numbers.
This result numerically checks the classic result for the steady
state problem, which states that the onset of the Rayleigh-B\'enard
instability does not depend on the Prandtl
number~\citep{Drazin81book}. This statement is recalled
expressing~\eqref{e:stab_eqn} in the
$(\za,\ta)$ space, re-scaling $\wsi$ and $\Tsi$ and their
derivatives to $\wai$ and $\Tai$ \emph{via}~\eqref{e:w_scale}, and
taking the limit when $\ts\to\infty$. 
The resulting equations are $\left(\pD{\za\za}-\ka^2\right)^2\wai = \ka^2\Tai$
and $\left(\pD{\za\za}-\ka^2\right)\Tai+\wai\Ra = 0$,
regardless of the value of Prandtl number, for which no assumption
has been made but to be positive. Now, as the resulting
expressions are only functions of $\za$, it is noted that the
latter equations also correspond to the linearized stability
system obtained assuming an exponential growth rate, $\exp\sigma\ta$, with a critical 
stability condition $\sigma=0$~\citep{Yang02pf}. This approach corresponds to the `marginal state'
variation of the frozen time model~\citep{Gresho71ijhmt}.
With this set of equations, the
computed critical Rayleigh number and its associated wavenumber
are $1707.7618$ and $3.11632$, respectively, in agreement with the
pair $(1707.765,3.12)$ proposed by \citet{Sparrow64jfm} and
$(1707.7618,3.11635)$, computed by \citet{Mizushima95jpsj}.

In numerical terms, the \TSI system assumption means that the outer boundary to
be considered goes to infinite. To reproduce this fact into the computation
of eigenvalues prior to the minimization process, the extrapolation procedure
described by~\citet{Chen83jfm} for critical $\Ras$ numbers was used. Roughly,
this approach is based on the observation that different $\Ras{}_n$ numbers,
obtained for different outer depths $\zs_n$, decrease approximately as a
geometrical sequence. Then, the asymptotic Rayleigh number can be computed as
$\Ras^{0} \approx \Ras{}_n^0 + r(\Ras{}_n^0-\Ras{}_{n-1}^0)/(1-r)$,
where $r$ is the slope of the approximately logarithmic line obtained
using different pairs $(\zs_n,\Ras{}_n)$.

\section{Results and discussion}\label{s:results_and_discussion}

\subsection{Base state solutions}\label{s:base_state}

The base state solution can be calculated using Laplace
transforms, which yields equation~\eqref{e:sol_lev}. This approach
has the advantage of producing a series with faster convergence
than that obtained through Fourier decomposition.
\begin{equation}
\Tso(\zs,\ts) = 1+\sum_{n\geq0}(-1)^{n+1}\left\{
        \erfc\left[\frac{n}{\sqrt{\ts}}+\frac{\zs}{2}\right] +
        \erfc\left[\frac{n+1}{\sqrt{\ts}}-\frac{\zs}{2}\right]\right\}\label{e:sol_lev}
\end{equation}
The \TSI solution can be readily obtained
solving~\eqref{e:a_base_state} on a semi-infinite domain:
\begin{equation}\label{e:sol_dp}
\Tso{}_{\TSI}(\zs) = \erf\left(\zs/2\right).
\end{equation}
For values of $\ts$ lower or close to $0.01$ very small relative
differences between equations~\eqref{e:sol_lev}
and~\eqref{e:sol_dp} are observed. Computing the latter as
$\lambda=\max_{\zs}\{100\times|1-\Tso/\Tso{}_{\TSI}|\}$, for 
$\ts=0.005$, $0.007$, $0.01$, $0.02$ and $0.05$, $\lambda<10^{-12}$, 
$10^{-12}$, $10^{-10}$, $10^{-4}$ and $0.1$, respectively.

\subsection{Comparison with the unsteady Rayleigh-B\'enard
problem}\label{s:comp_RB} 
The deduction of the nonpenetrative stability problem in the light of 
the propagation model yields an interesting similitude with the unsteady 
Rayleigh-B\'enard problem studied by~\citet{Kim99ijhmt}. Eqs.~\eqref{e:stab_eqn1}
and~\eqref{e:stab_eqn2} have the same analytical expression than
those corresponding to the latter work. The only difference
between them is the thermal condition imposed at the boundary away
from the step change in temperature (named herein as the outer
boundary). In the present problem, the boundary condition
$\lim_{\zs\to\infty}\DD\Tsi=0$ is imposed, while
in~\citet{Kim99ijhmt}, $\lim_{\zs\to\infty}\Tsi=0$ is imposed
instead, representing the existence of an isothermal outer
boundary. It can be shown, however, that both types of outer
boundary condition must be satisfied simultaneously in both
problems.
For fixed Rayleigh and Prandtl numbers, an onset time $\tc$ exists
such that $\Tsi=0$ for $\ts<\tc=\tc(\Pra,\Ra)$ (i.e., the system
does not experience convection before the onset time). On the
other hand, the similarity condition inherent to the present
propagation model imposes the scaling $\delta_{\theta}\sim\sqrt{\ts}\ll1$
($\ts\lll1$) for the thermal penetration depth, and the boundary
condition $\DD\Tsi(1/\sqrt{\ts})=0$. Then, considering, as stated
previously, that the perturbations are mainly confined within the
TBL, and assuming continuity of the temperature disturbance,
necessarily $\Tsi(1/\sqrt{\ts})=0$, which is a mere consequence of
the small penetration depth occurring at small times. As this
holds for arbitrarily small onset times (and, consequently,
arbitrarily large values of the Rayleigh number), if
$\lim_{\zs\to\infty}\DD\Tsi=0$, then $\lim_{\zs\to\infty}\Tsi=0$.
Consequently, both problems, the impulsively isothermally heated
Rayleigh-B\'enard and the present nonpenetrative convection, are
equivalent, provided the existence of a Rayleigh number range that
support the deep-pool assumption.

Another interesting feature of the
eigensystem~\eqref{e:stab_eqn} is that its
eigenvalues are insensitive to the type of outer boundary
condition considered, free or rigid, as can be verified with
arguments similar to those of the previous paragraph. Hence,
according to the propagation model, for high Rayleigh numbers the
only boundary that matters to eigenvalues (both in the thermal and
kinematic sense) is the one subjected to the impulsive change on
temperature. This conclusion agrees with that
of~\citet{Foster65b_pf}, who noticed that motion was `decoupled
from the bottom', analyzing the problem of a surface-stress-free
fluid layer subject to a step change in temperature, using the
amplification model. Another consequence of this conclusion is
that the free-rigid results to be presented here should be valid
for the free-free and free-rigid variations of the Rayleigh-B\'enard
convection. The same applies, of course, to the free-free
nonpenetrative convection problem, which offers a reasonable
approximation to systems where a nearly stress-free, strong and
stable density interface exists between two rather homogeneous
layers of fluid. The latter, so-called
`thermocline'~\citep{Imberger90aam}, is commonly found in lakes
and reservoirs.

\subsection{Solution of the eigenvalue problem}

As $\ts\leq0.01$ (which is bonded to the assumption of a highly
supercritical system) must hold to keep the self similarity of the
base state, a lower bound to valid Rayleigh numbers is imposed:
\begin{equation}\label{e:min_Ra}
\Ra(\Pra) \geq
\Ra_{\text{min}}(\Pra)=\frac{\Ras(\Pra)}{0.01^{3/2}}
\end{equation}
Despite the existence of~\citet{Kim99ijhmt} results for the
impulsively heated Rayleigh-B\'enard problem, whose mathematical
posing fully coincides with that of the present nonpenetrative
convection problem as discussed in previous section, the
eigenvalues for the rigid-rigid case were re-calculated here to
serve as an additional validation of the numerical results
obtained. Differences on computed values of $\Ras$ were found only
for $\Pra=100$, and even in that case they were not higher than
about 1\%. In the case of the $\ks$ computed values, a difference
(of about 10\%) was found for $\Pra=1$. The latter are indicative
of an apparent error on Kim et al.'s~\citep{Kim99ijhmt} solution.
Table~\ref{t:stab_data} shows the minimum $(\ks, \Ras)$
eigenvalues computed for the \TSI system for a range of Prandtl
numbers and the corresponding values of the minimum valid Rayleigh
number, given by~\eqref{e:min_Ra}.

\begin{table}[!h]
\caption{Critical $(\ks,\Ras)$ parameters found for the \TSI
system, as a function of Prandtl number. The fourth and fifth
columns (labelled as KCC99) show the critical numbers found for
the transient Rayleigh-B\'enard problem studied
by~\citet{Kim99ijhmt}. Columns $6$ and $9$ show the minimum
Rayleigh numbers that guarantee that the \TSI assumption is valid
both for the rigid-rigid and for the free-rigid system,
respectively.}\label{t:stab_data}
\begin{center}
\begin{tabular}{lcccccccc}\toprule
 & \multicolumn{5}{c}{Rigid-rigid} & \multicolumn{3}{c}{Free-rigid}\\\cmidrule{2-6}
 & \multicolumn{2}{c}{Present work} & \multicolumn{2}{c}{KCC99} & & \\\midrule
$\Pra$  & $\ks$     & $\Ras$    & $\ks$ & $\Ras$    &
$\Ra_{\text{min}}$    & $\ks$     & $\Ras$    & $\Ra_{\text{min}}$
\\\midrule
$0.01$  & $0.824$   & $1799.06$ & $0.82$& $1799.1$  & $1.80\times10^6$  & $0.809$   & $1675.92$ & $1.68\times10^6$  \\
$0.1$   & $0.813$   & $219.10$  & $0.81$& $219.1$   & $2.19\times10^5$  & $0.766$   & $180.98$  & $1.81\times10^5$  \\
$0.71$  & $0.725$   & $53.56$   & ---   & ---       & $5.36\times10^4$  & $0.637$   & $36.58$   & $3.66\times10^4$  \\
$1$     & $0.702$   & $44.81$   & $0.63$& $44.81$   & $4.48\times10^4$  & $0.607$   & $29.36$   & $2.94\times10^4$  \\
$7$     & $0.589$   & $24.73$   & ---   & ---       & $2.47\times10^4$  & $0.447$   & $12.68$   & $1.27\times10^4$  \\
$100$   & $0.538$   & $20.97$   & $0.54$& $20.70$   & $2.10\times10^4$  & $0.337$   & $9.01$    & $9.01\times10^3$  \\
$1000$  & $0.533$   & $20.70$   & $0.53$& $20.69$   & $2.07\times10^4$  & $0.320$   & $8.67$    & $8.67\times10^3$  \\
$\infty$& $0.533$   & $20.67$   & $0.53$& $20.67$   &
$2.07\times10^4$  & $0.317$   & $8.63$    & $8.63\times10^3$
\\\bottomrule
\end{tabular}    
\end{center}

\end{table}

\begin{figure}[!h]
\begin{center}
\resizebox{\textwidth}{!}{%
\includegraphics{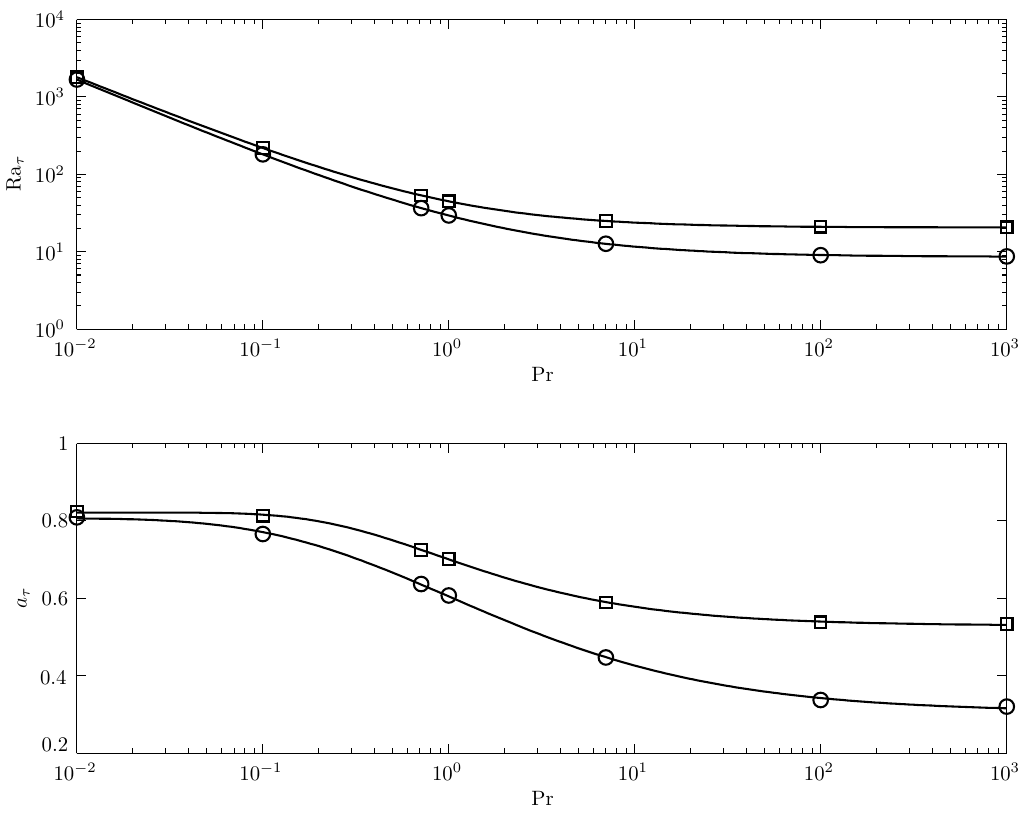}}
\caption{Upper panel: effect of the Prandtl number on $\Ras$ for
the rigid-rigid and rigid-free cases. Lower panel: effect of the
Prandtl number on the wavenumber of the fastest growing horizontal
mode, \kac. Symbols represent calculated points, corresponding to
Table~\ref{t:stab_data}: squares for the rigid-rigid case and
circles for the free-rigid case. Curves represent interpolated
results using the models given by Eqs.~\eqref{e:corr_dp_Ra},
and~\eqref{e:corr_dp_a} and parameter sets given by
Table~\ref{t:ts_params}.}\label{f:Ra_Pr}
\end{center}
\end{figure}

Results for the \TSI free-rigid case, which have not been
previously reported in the context of the present problem and
method, are shown in Fig.~\ref{f:Ra_Pr}. The $\Ras$ parameter
varies exponentially for $\Pra\leq1$ and is virtually constant for
values of $\Pra>1000$. The same trend occurs for the rigid-rigid
case, as previously commented by~\citet{Kim99ijhmt,Kim02tcfd}.
Considering comments on Section~\ref{s:comp_RB}, present data
extend the results reported in the former work and are new to the
nonpenetrative problem context. The following correlations, valid
for $0.01\leq\Pra\leq1000$, can be used to predict the onset time
and the most unstable mode in the case of the \TSI system, for the
free-rigid or rigid-rigid cases, with an error bound of 2\%:
\begin{gather}
\ts{}_c = a_1\left[a_2+\left(\frac{a_3}{\Pra}\right)^{a_4}\right]^{a_5}\Ra^{-2/3}\label{e:corr_dp_Ra}\\
\kac = \left(b_1 + b_2\erf\left[\left(\frac{b_3}{\Pra} \right)^{b_4}\right]^{b_5}\right)\ts{}_c^{-1/2}\label{e:corr_dp_a}
\end{gather}
Corresponding values of the parameters $a_j$ are given on Table~\ref{t:ts_params}.
For higher values of the
Prandtl number ($\Pra>1000$), $\ts{}_c = a_{\infty}\Ra^{-2/3}$ and 
$\kac    = b_{\infty}\ts{}_c^{-1/2}$ replace~\eqref{e:corr_dp_Ra} 
and~\eqref{e:corr_dp_a}, respectively. Here, $a_{\infty}=7.531$ 
and $4.207$, $b_{\infty}=0.533$ and $0.317$, for the rigid-rigid 
and free-rigid cases, respectively.

\begin{table}[!h]
\caption{Parameters $a_j$ and $b_j$ of Eqs.~\eqref{e:corr_dp_Ra}
and~\eqref{e:corr_dp_a} for the rigid-rigid (RR) and free-rigid
(FR) conditions.}\label{t:ts_params}
\begin{center}
\begin{tabular}{ccccc}\toprule
 & \multicolumn{2}{c}{$a_j$} & \multicolumn{2}{c}{$b_j$}\\\cmidrule{2-5}
 $j$  & RR         & FR          & RR       & FR\\\midrule
 $1$  & $9.8371$   & $5.9017$    & $0.5291$ & $0.3066$\\
 $2$  & $1.9022$   & $1.3279$    & $0.2923$ & $0.5002$\\
 $3$  & $2.0867$   & $2.5505$    & $0.6329$ & $1.1196$\\
 $4$  & $0.8502$   & $0.7730$    & $0.3347$ & $0.1971$\\
 $5$  & $1.1421$   & $1.3132$    & $2.0924$ & $3.2267$\\\bottomrule
\end{tabular}    
\end{center}
\end{table}

\begin{figure}[!h]
\begin{center}
\resizebox{\textwidth}{!}{%
\includegraphics{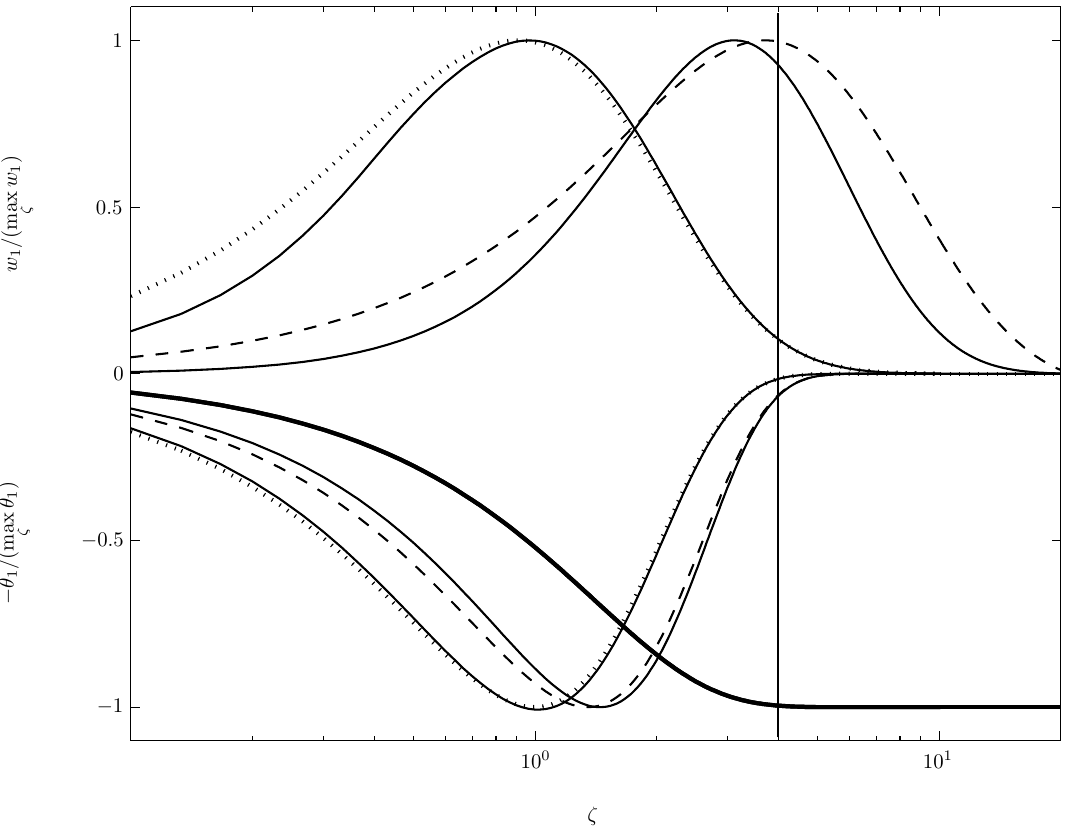}}
\caption{Normalized amplitude functions for (minus) temperature
and vertical velocity disturbances (lower and upper half,
respectively). Solid lines represent the rigid-rigid case in both
sets. From left to right, the latter curves show computed results
for $\Pra=0.01$ and $\Pra\to\infty$. The dotted and dashed curves
represent the free-rigid case for $\Pra=0.01$ and $\Pra\to\infty$,
respectively. The bold monotonic curve exhibits (minus) the base
state temperature, given by~\eqref{e:sol_dp}.}\label{f:amp_tsi}
\end{center}
\end{figure}

The amplitude functions corresponding to the results in
Table~\ref{t:stab_data}
 are represented in Fig.~\ref{f:amp_tsi}.
Here, a TBL can be defined as the $\zs$ value for which the base
temperature reaches a value of $0.99$. This limiting condition is
depicted in Fig.~\ref{f:amp_tsi} with a vertical line. With
this definition, the latter figure shows a tendency of the
amplitude curves
 to displace out of the TBL with increasing Prandtl number.
The same trend was previously observed by~\citet{Kang97pf} in the
\TSI system associated with the B\'enard-Marangoni convection.

For Prandtl numbers greater than about one, it is found that
vertical velocity disturbances reach depths that exceed by a
factor close to $2$ the thermal penetration depth. The increasing
of the penetration of disturbances with Prandtl number means that
the higher the latter parameter, the deeper is the layer where
disturbances exist (Fig.~\ref{f:amp_tsi}). At the same time, as
the Prandtl number increases, the system becomes less stable as
shown by the monotonically decreasing marginal stability curves of
the upper panel of Fig.~\ref{f:Ra_Pr}.

An interesting feature of the eigenfunctions is that only for
medium to large Prandtl numbers (greater that about $10$) the
asymptotic decay of the disturbances with depth in the case of the
free-rigid case is noticeably slower than in the rigid-rigid case
(Fig.~\ref{f:amp_tsi}). This trend is consistent with the
separation between the stability curves for different boundary
conditions depicted in Fig.~\ref{f:Ra_Pr} (upper panel), which
appears to be minimum for low Prandtl numbers and maximum in the
infinite Prandtl number case.
As some liquid metals, like mercury, have very low Prandtl numbers
($\sim0.025$ at room temperature), present results suggest a way
to avoid the kinematic effect of the boundary condition in
laboratory experiments with a proper choice of the fluid.

Computed critical wavenumbers exhibit slight variations with the
Prandtl number, for values of this parameter lower than about
$0.1$ and larger than about $10$ (Fig.~\ref{f:Ra_Pr}, lower
panel). In the intermediate range ($0.1 < \Pra < 10$), however,
they present rather steep change rates with \Pra. As the onset of
convection is marked by the formation of regular cells, within the
intermediate Pr range those disturbances should be rather more
sensitive to small spatial variations in fluid properties than in
the low and high Pr cases. Consequently, it is believed that this
factor may influence to some extent the reproducibility of
experiments and possibly explain in part the large dispersion in
the horizontal wavelengths experimentally obtained
by~\citet[][Fig.~4]{Foster69pf}.

Present results have some differences with respect to previous
numerical calculations using other other approaches to the
stability analysis. 
In particular, in the amplification
model~\citep{Foster65b_pf} an amplification factor, built upon the
normalized RMS of the disturbance of the vertical velocity field,
is defined as 
$\bar{w}(t) = 
[\int_0^1\wi^2(\za,\ta)\,\dd\za/\int_0^1\wi^2(\za,0)
	\,\dd\za]^{1/2}$,
where $\wi(\za,0)$ represents the initial disturbance condition,
which has been commonly chosen as white noise with equal amplitude
coefficients \citep[see][]{Foster65b_pf,Mahler68pf,Gresho71ijhmt}.
When $\bar{w}(t)$ grows beyond some predefined factor, the
corresponding time is marked as the onset time. In this context,
different thresholds for $\bar{w}$ induce the estimation of
different times. Such need for a definition of limiting conditions
precludes a straightforward comparison between results coming from
different methods and care should be taken. The onset time
predicted by the present method is analyzed in more detail next.

\subsection{Analysis of onset time}

To assess the onset time, commonly three classes of characteristic
times are considered. The first corresponds to that which comes
indirectly from the
eigensystem~\eqref{e:stab_eqn1}--\eqref{e:stab_eqn2}, $\tc$. The
second one is that which marks the thermal dominance of advection
over diffusion, $\tu$. Finally, the third one is that at which
fluid motion or temperature increase can be experimentally
detected, $\tm$. It is likely that the better the experiment, the
closer is $\tm$ to $\tu$ since normally scalar change sensing is
used. Experimental verification of $\tc$ seems to be more
difficult, as it marks the beginning of convection, with a very
small amplitude fluid
motion~\citep{Foster69pf,Davenport74ijhmt,Yang02pf,Chung04kjce,Choi04ijts}.
On the other hand, there must be a lapse of time when velocities
are small enough to make the advective term in the energy
equation negligible compared with the diffusive
one~\citep{Elder69jfm}, that is
$0<\wdi\pD{\zd}\Ta\ll\alpha\laplacea\Ta$
for $\tas$ such that
$\tc<\tas<\tu$. Then, $\tc$ must always be lower that $\tu$.

In the case of results for the rigid-rigid
case,~\citet{Kim99ijhmt} supported the conjecture
of~\citet{Foster69pf} about the existence of a scaling factor of
about $4$, between time $\tc$ coming from eigenvalue calculations
and time $\tm$ corresponding to observations of convective motion.
To this purpose, they used the propagation model with the
temperature step change setup and compared their theoretical
results with the experimental ones by~\citet{Ueda84pf}. Further
comparisons were later reported by the same research
group~\citep[see][and references therein]{Kim02tcfd,Choi04ijts}
for large Prandtl numbers.

For the free-rigid case, theoretical results using the
amplification model are given by~\citet{Foster65b_pf} for step and
ramp changes in surface temperature with an isothermal bottom and
some $\Ra$-$\Pra$ combinations. However, no experimental results
were available to validate the former case. Also in a theoretical
framework, for $\Pra=7$ and free-free conditions, defining the
onset time, $\tu$, from a Nusselt number departure of $1$\% above
the conductive state,~\citet{Jhaveri82jfm}, found that
$\tu\sim\Ra^{-2/3}$ (named hereafter as the $-2/3$ power law), and
also that $\kac\sim\Ra^{1/3}$, showing that the latter relations
hold for $\Ra\geq30\times(27/4)\pi^4\approx2\times10^4$, which is
close to the corresponding lower limit of this parameter proposed
in Table~\ref{t:stab_data}. From their data and onset time
 definition (considered herein as being representative of \tu),
the value $\Ra\,\tu^{3/2}\approx350$ is obtained. This value is
greater than the critical $\Ras=12.68=\Ra\,\tc^{3/2}$ obtained
from Table~\ref{t:stab_data}, thus giving a value of the ratio
$\tu/\tc=(350/12.68)^{2/3}\approx9.1$.

\begin{figure}[!h]
\begin{center}
\resizebox{\textwidth}{!}{%
\includegraphics{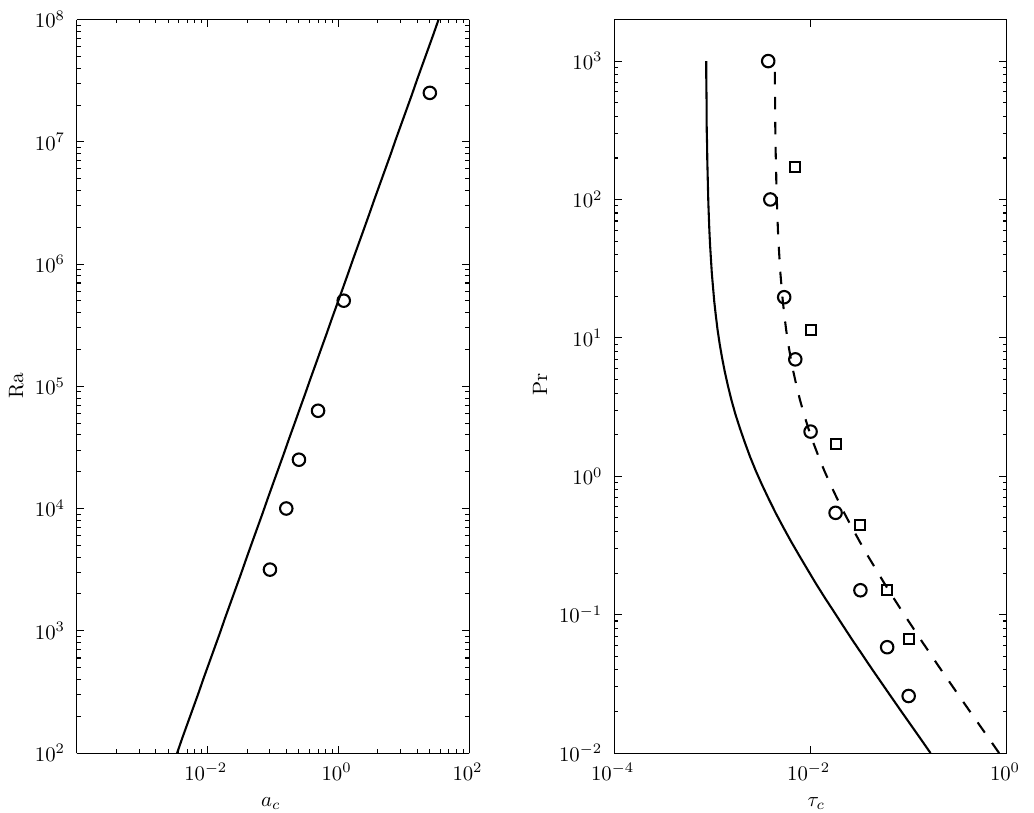}}
\caption{Left panel: comparison between critical wavenumbers
computed using the present propagation model (solid line) and the
ones reported by~\citet{Foster65b_pf} (circles), for the
free-rigid case, step change in temperature, with $\Pra=7$. Right
panel: Critical time as a function of Prandtl number; comparison
between results obtained using propagation and amplification
models with $\Ra=10^6$. Solid
 line represents results obtained with the present propagation model, while the
 dashed line shows the latter predictions amplified by a factor of  $5$.
Circles represent results from~\citet{Foster65b_pf},
 using an amplification factor $\bar{w}=10$; squares represent
equivalent results with $\bar{w}=100$.}\label{f:Foster_chk_fr}
\end{center}
\end{figure}

In Fig.~\ref{f:Foster_chk_fr} (right panel), results from both the
present propagation theory and the amplification
model~(\citet{Foster65b_pf}) for different Prandtl numbers and
$\Ra=10^6$ are shown.
 It can be seen that for the free-rigid
case both models differ on computed onset times by a factor close
to $5$, when $\bar{w}=10$ is used as an amplification factor, and
the Prandtl number is greater than about $10$. For lower values of
\Pra, the propagation model yields higher onset times than the
ones found using the amplification model. It is noteworthy that
the best amplification ratios for the rigid-rigid experiment
by~\citet{Foster69pf} were found between $\bar{w}=10^3$ and $10^8$
(the latter theoretical calculations were previously reported
by~\citet{Foster68pf}). Taking these results into account, it can
be concluded that the tuned amplification factor can also be
understood as a measure of the disturbance level that a system can
afford just before the onset of convection. Wavenumbers were
calculated with the present model using the data for $\Pra=7$
shown in Table~\ref{t:stab_data}, i.e.,
$\ks=0.447=\kac\,\sqrt{\tc}$. Fig.~\ref{f:Foster_chk_fr} (left
panel) shows good agreement between present computations of the
critical wavenumber $\kac$ and those of~\citet{Foster65b_pf} for
Rayleigh numbers higher than about $10^5$, verifying the $1/3$
power law scaling previously noted.

Experimental data for the free-rigid unsteady Rayleigh-B\'enard case
have been reported by~\citet{Spangenberg61pf}
and~\citet{Foster65pf}, but a step in surface temperature was not
obtained since evaporative cooling was the dominant effect. An
approximate piecewise linear cooling on top was found
experimentally. Using infrared radiometry to record surface layer
temperature, the latter author compared his results with those
obtained by the former, checking them against calculations made
with the amplification model~\citep{Foster65b_pf}. Good
correspondence with this theory was found, however, due to the
linear evolution of temperature on the top boundary, the
experimental results fitted better a $-2/5$ exponent, instead of
the $-2/3$ power law expected for the step change case. Table~I
of~\citet{Foster65pf} lists several results for the onset time
given combinations of Prandtl and Rayleigh numbers. That table
also includes results from~\citet{Spangenberg61pf}. This data set
agree well with amplification model calculations
by~\citet{Foster65b_pf} using amplification factors between $10$
and $100$, showing that computed values of the onset times for low
amplification and linear cooling describe well the onset of
evaporative convection, as previously mentioned. On the other
hand, the latter measurements yield times that differ in
approximately three orders of magnitude with the present
propagation theory results.
Differences appear to reside solely on the different applicable
power laws, with
$\tc{}_{\,\text{step}}/\tc{}_{\,\text{linear}}\sim10^{-3}$ in the
range of $\Ra$ values analyzed, since for the step cooled system
$\Ra\,\tc^{3/2}=C_1$ (constant), while in the linearly cooled one
the scaling is rather $\Ra\,\tc^{5/2}=C_2$ (constant).

In the case of convection induced by gas absorption with free-rigid
boundaries, defining a Rayleigh number based on a concentration step, 
$\Ra=g\gamma(C-C_b)L^3 D^{-1}\nu^{-1}$ ($\gamma$, $C$, $C_b$ and $D$ 
are the concentration coefficient of expansion, equilibrium and bulk 
concentration of solute and mass diffusion coefficient, respectively), 
along with the time scale $L^2/D$, Plevan and Quinn's 
data~\citep{Plevan66aj} yield measured dimensionless onset times 
$\tm\approx2.1\times10^{-3}$ for carbon dioxide in water and 
$\tm\approx1.3\times10^{-4}$ for sulphur dioxide in water.
Although $\Pra\approx6.25$ in both cases, differences may come
partly from the better solubility of the latter gas in water~\citep{Blair69jfm}. Corresponding 
time ratios, compared with that obtained from Eq.~\eqref{e:corr_dp_Ra}, 
are $\tm/\tc\approx11$ and $13.1$, respectively. 
Similarly,~\citet{Blair69jfm} found  $\Ra\,\tm^{3/2}\approx300$
for sulphur dioxide in water. Using data from
Table~\ref{t:stab_data}, time ratios are 
$\tm/\tc\approx8.2$, $10.3$ and
$10.6$, for Prandtl numbers of $7$, $100$, and $1\,000$, respectively.
In both works, $\Ra\gtrsim10^{6}$.
Unfortunately, it is impossible to build similar relations with the 
experimental setup information from~\citet{Tan92ces}, since no liquid layer 
thickness was specified in that paper. On the other hand, in the latter
work, an alternative temporal and depth-dependent version of 
the Rayleigh number is proposed, along with a theoretical model 
where the solution of 
$\Ra(\zd,\td)=\zd{}^4 g \gamma \mu^{-1}D^{-1}\mathrm{d}C/\mathrm{d}\zd$
is maximized with respect to $\zd$, finding the length scale $L(\td) = 2\sqrt{2D\td}$. 
The corresponding onset time is computed using the 
critical Rayleigh number in B\'enard convection on a steady, horizontally infinite
domain with free-rigid boundaries~\citep{Chandrasekhar61book} on the
expression for the maximum $\Ra(\zd,\td)$. From~\citet{Tan92ces} data, 
the latter was found to be on the order of $1\,000$.
Using $\Ra(\zd,\td)$ and $L(\td)$ as 
the corresponding Rayleigh number and length scale, yields 
time ratios $\tm/\tc$ close to $4$, but their definitions are not analogous to the
present Rayleigh number
and length scale. 
Consequently, except for Tan and Thorpe's data~\citep{Tan92ces}, which provides no clue, 
all the revised references support the present estimation of a time ratio $\tm/\tc$ on the 
order of $10$, rather that $4$, for the free-rigid system.

\section{Concluding remarks}

A stability analysis using propagation theory has been conducted
to predict the factors that rule the temporal dependence of the
onset of nonpenetrative convection in an initially isothermal
Boussinesq fluid. It was shown that for the \TSI system, or, in
other words, for high thermal disturbances, which are defined in
terms of Rayleigh numbers that exceed a certain minimum for given
Prandtl numbers (Table~\ref{t:stab_data}), the study of
nonpenetrative convection equates conceptually and numerically the
unsteady Rayleigh-B\'enard convection. An extension of previously
reported results for the rigid-rigid system using propagation
theory~\citep{Kim99ijhmt} has been proposed for free-rigid
boundary conditions. As several works have reported the study of
the onset of unsteady Rayleigh-B\'enard convection, a comparison of
their results obtained with different methods, with those obtained
using the present linear model was made. For Rayleigh and Prandtl
numbers within the limits of the present theory, good agreement
was found between present results and theoretical ones obtained
with the amplification model~\citep{Foster65b_pf} and the
stochastic method~\citep{Jhaveri82jfm}.  General agreement on the
validity of the scaling $\tc\sim\Ra^{-2/3}$ was found. On the
other hand, numerical evidence along with experimental data,
suggest that the lag between theoretical onset times ($\tc$) and
detected ones ($\tu$ or $\tm$) is dominated at least by two conditions,
namely, the kinematic boundary condition on the side where the
heat flow (or temperature change) is imposed, and the way heating 
(or cooling) is applied in time. It is argued that, at difference
from the theoretical determination of $\tc$ or $\tu$, recording of
$\tm$ depends in great extent on the experiment configuration and
technological limitations. In particular, present results suggest
that for medium to large Prandtl numbers (greater than about $1$)
and a top (if cooled from above) stress-free boundary, an onset
time relation of $\tm/\tc\sim10$ rather than $4$ (previously
proposed for the rigid-rigid system), seems to fit the available
data reasonably well. Given these results, it is concluded that
the latter values of the $\tm/\tc$ ratio are particular cases of a
more complex function that should take into account, at least,
boundary conditions for the prediction of the
onset of convective motion from the experimental knowledge of
changes in scalar fields.

\section*{Acknowledgements}

The authors gratefully acknowledge support from the Chilean
National Commission for Scientific and Technological Research,
CONICYT, the Department of Civil Engineering of the University of
Chile, and Fondecyt Project No. 1040494.


\begin{thebibliography}{37}
\expandafter\ifx\csname natexlab\endcsname\relax\def\natexlab#1{#1}\fi
\providecommand{\url}[1]{\texttt{#1}}
\providecommand{\href}[2]{#2}
\providecommand{\path}[1]{#1}
\providecommand{\DOIprefix}{doi:}
\providecommand{\ArXivprefix}{arXiv:}
\providecommand{\URLprefix}{URL: }
\providecommand{\Pubmedprefix}{pmid:}
\providecommand{\doi}[1]{\href{http://dx.doi.org/#1}{\path{#1}}}
\providecommand{\Pubmed}[1]{\href{pmid:#1}{\path{#1}}}
\providecommand{\bibinfo}[2]{#2}
\ifx\xfnm\relax \def\xfnm[#1]{\unskip,\space#1}\fi
\bibitem[{Adrian(1986)}]{Adrian86ef}
\bibinfo{author}{Adrian, R.J.}, \bibinfo{year}{1986}.
\newblock \bibinfo{title}{Turbulent thermal convection in wide horizontal fluid layers}.
\newblock \bibinfo{journal}{Exp. Fluids} \bibinfo{volume}{4}, \bibinfo{pages}{121}.
\bibitem[{Blair and Quinn(1969)}]{Blair69jfm}
\bibinfo{author}{Blair, L.M.}, \bibinfo{author}{Quinn, J.A.}, \bibinfo{year}{1969}.
\newblock \bibinfo{title}{Onset of cellular convection in a fluid layer with time-dependent density gradients, the}.
\newblock \bibinfo{journal}{J. Fluid Mech.} \bibinfo{volume}{36}, \bibinfo{pages}{385--400}.
\bibitem[{Chandrasekhar(1961)}]{Chandrasekhar61book}
\bibinfo{author}{Chandrasekhar, S.}, \bibinfo{year}{1961}.
\newblock \bibinfo{title}{Hydrodynamic and hydromagnetic Stability}.
\newblock \bibinfo{edition}{First} ed., \bibinfo{publisher}{Oxford: Clarendon}.
\bibitem[{Chen et~al.(1983)Chen, Chen and Sohn}]{Chen83jfm}
\bibinfo{author}{Chen, K.}, \bibinfo{author}{Chen, M.M.}, \bibinfo{author}{Sohn, C.W.}, \bibinfo{year}{1983}.
\newblock \bibinfo{title}{Thermal instability of two-dimensional stagnation-point boundary layers}.
\newblock \bibinfo{journal}{J. Fluid Mech.} \bibinfo{volume}{132}, \bibinfo{pages}{49--63}.
\bibitem[{Choi(2004)}]{Choi04ijhmt}
\bibinfo{author}{Choi, C.K.}, \bibinfo{year}{2004}.
\newblock \bibinfo{title}{The onset of convective instability in a horizontal fluid layer subjected to a constant heat flux from below}.
\newblock \bibinfo{journal}{Int. J. Heat Mass Transfer} \bibinfo{volume}{47}, \bibinfo{pages}{4377--4384}.
\bibitem[{Choi et~al.(1988)Choi, Lee, Hwang and Yoo}]{Choi88proc}
\bibinfo{author}{Choi, C.K.}, \bibinfo{author}{Lee, J.D.}, \bibinfo{author}{Hwang, S.T.}, \bibinfo{author}{Yoo, J.S.}, \bibinfo{year}{1988}.
\newblock \bibinfo{title}{The analysis of thermal instability and heat transfer prediction in a horizontal fluid layer heated from below}.
\newblock \bibinfo{journal}{Proceedings of the International Conference On Fluid Mechanics} , \bibinfo{pages}{1193--1198}.
\bibitem[{Choi et~al.(2004a)Choi, Park and Kim}]{Choi04hmt}
\bibinfo{author}{Choi, C.K.}, \bibinfo{author}{Park, J.H.}, \bibinfo{author}{Kim, M.C.}, \bibinfo{year}{2004}a.
\newblock \bibinfo{title}{The onset of buoyancy-driven convection in a horizontal fluid layer subjected to evaporative cooling}.
\newblock \bibinfo{journal}{Heat Mass Transfer} \bibinfo{volume}{41}, \bibinfo{pages}{155--162}.
\bibitem[{Choi et~al.(2004b)Choi, Park, Park, Cho, Chung and Kim}]{Choi04ijts}
\bibinfo{author}{Choi, C.K.}, \bibinfo{author}{Park, J.H.}, \bibinfo{author}{Park, H.K.}, \bibinfo{author}{Cho, H.J.}, \bibinfo{author}{Chung, T.J.}, \bibinfo{author}{Kim, M.C.}, \bibinfo{year}{2004}b.
\newblock \bibinfo{title}{Temporal evolution of thermal convection in an initially stably-stratified horizontal fluid layer}.
\newblock \bibinfo{journal}{Int. J. Therm. Sci.} \bibinfo{volume}{43}, \bibinfo{pages}{817--823}.
\bibitem[{Chung et~al.(2004)Chung, Kim and Choi}]{Chung04kjce}
\bibinfo{author}{Chung, T.J.}, \bibinfo{author}{Kim, M.C.}, \bibinfo{author}{Choi, C.K.}, \bibinfo{year}{2004}.
\newblock \bibinfo{title}{Temporal evolution of thermal instability in fluid layers isothermally heated from below, the}.
\newblock \bibinfo{journal}{Korean J. Chem. Eng.} \bibinfo{volume}{21}, \bibinfo{pages}{41--47}.
\bibitem[{Davenport and King(1974)}]{Davenport74ijhmt}
\bibinfo{author}{Davenport, I.F.}, \bibinfo{author}{King, C.J.}, \bibinfo{year}{1974}.
\newblock \bibinfo{title}{The onset of natural convection from time-dependent profiles}.
\newblock \bibinfo{journal}{Int. J. Heat Mass Transfer} \bibinfo{volume}{17}, \bibinfo{pages}{69--76}.
\bibitem[{Drazin and Reid(1981)}]{Drazin81book}
\bibinfo{author}{Drazin, P.G.}, \bibinfo{author}{Reid, W.H.}, \bibinfo{year}{1981}.
\newblock \bibinfo{title}{Hydrodynamic Stability}.
\newblock \bibinfo{publisher}{Cambridge University Press}.
\bibitem[{Elder(1969)}]{Elder69jfm}
\bibinfo{author}{Elder, J.W.}, \bibinfo{year}{1969}.
\newblock \bibinfo{title}{Temporal development of a model of high {Rayleigh} number convection, the}.
\newblock \bibinfo{journal}{J. Fluid Mech.} \bibinfo{volume}{35}, \bibinfo{pages}{417--437}.
\bibitem[{Foster(1965a)}]{Foster65pf}
\bibinfo{author}{Foster, T.D.}, \bibinfo{year}{1965}a.
\newblock \bibinfo{title}{Onset of convection in a layer of fluid cooled from above}.
\newblock \bibinfo{journal}{Phys. Fluids} \bibinfo{volume}{8}, \bibinfo{pages}{1770--3}.
\bibitem[{Foster(1965b)}]{Foster65b_pf}
\bibinfo{author}{Foster, T.D.}, \bibinfo{year}{1965}b.
\newblock \bibinfo{title}{Stability of homogeneous fluid cooled from above}.
\newblock \bibinfo{journal}{Phys. Fluids} , \bibinfo{pages}{1249--1257}.
\bibitem[{Foster(1968)}]{Foster68pf}
\bibinfo{author}{Foster, T.D.}, \bibinfo{year}{1968}.
\newblock \bibinfo{title}{Effect of boundary conditions on the onset of convection}.
\newblock \bibinfo{journal}{Phys. Fluids} \bibinfo{volume}{11}, \bibinfo{pages}{1257}.
\bibitem[{Foster(1969)}]{Foster69pf}
\bibinfo{author}{Foster, T.D.}, \bibinfo{year}{1969}.
\newblock \bibinfo{title}{Onset of manifest convection in a layer of fluid with a time-dependent surface temperature}.
\newblock \bibinfo{journal}{Phys. Fluids} \bibinfo{volume}{12}, \bibinfo{pages}{2482--2487}.
\bibitem[{Goldstein and Volino(1995)}]{Goldstein95jhtta}
\bibinfo{author}{Goldstein, R.J.}, \bibinfo{author}{Volino, R.J.}, \bibinfo{year}{1995}.
\newblock \bibinfo{title}{Onset and development of natural convection above a suddenly heated surface}.
\newblock \bibinfo{journal}{J. Heat Trans.-T. Asme} \bibinfo{volume}{117}, \bibinfo{pages}{808}.
\bibitem[{Gresho and Sani(1971)}]{Gresho71ijhmt}
\bibinfo{author}{Gresho, P.M.}, \bibinfo{author}{Sani, R.L.}, \bibinfo{year}{1971}.
\newblock \bibinfo{title}{Stability of a fluid layer subjected to a step change in temperature: Transient vs. frozen time analysis, the}.
\newblock \bibinfo{journal}{Int. J. Heat Mass Transfer} \bibinfo{volume}{14}, \bibinfo{pages}{207--221}.
\bibitem[{Homsy(1973)}]{Homsy73jfm}
\bibinfo{author}{Homsy, G.M.}, \bibinfo{year}{1973}.
\newblock \bibinfo{title}{Global stability of time-dependent flows: Impulsively heated or cooled fluid layers}.
\newblock \bibinfo{journal}{J. Fluid Mech.} \bibinfo{volume}{60}, \bibinfo{pages}{129}.
\bibitem[{Ihle and Ni\~{n}o(2005)}]{Ihle05xxx}
\bibinfo{author}{Ihle, C.F.}, \bibinfo{author}{Ni\~{n}o, Y.}, \bibinfo{year}{2005}.
\newblock \bibinfo{title}{Global stability of nonpenetrative convection}.
\newblock \bibinfo{journal}{In preparation} .
\bibitem[{Imberger and Patterson(1990)}]{Imberger90aam}
\bibinfo{author}{Imberger, J.}, \bibinfo{author}{Patterson, J.}, \bibinfo{year}{1990}.
\newblock \bibinfo{title}{Physical limnology}.
\newblock \bibinfo{journal}{Adv. in Applied Mechanics} \bibinfo{volume}{27}, \bibinfo{pages}{303--475}.
\bibitem[{Jhaveri and Homsy(1982)}]{Jhaveri82jfm}
\bibinfo{author}{Jhaveri, B.S.}, \bibinfo{author}{Homsy, G.M.}, \bibinfo{year}{1982}.
\newblock \bibinfo{title}{The onset of convection in fluid layers heated rapidly in a time-dependent manner}.
\newblock \bibinfo{journal}{J. Fluid Mech.} \bibinfo{volume}{114}, \bibinfo{pages}{251--260}.
\bibitem[{Kang and Choi(1997)}]{Kang97pf}
\bibinfo{author}{Kang, K.H.}, \bibinfo{author}{Choi, C.K.}, \bibinfo{year}{1997}.
\newblock \bibinfo{title}{A theoretical analysis of the onset of surface-tension-driven convection in a horizontal liquid layer cooled suddenly from above}.
\newblock \bibinfo{journal}{Phys. Fluids} \bibinfo{volume}{9}, \bibinfo{pages}{7--15}.
\bibitem[{Kim and Kim(1986)}]{Kim86ijhmt}
\bibinfo{author}{Kim, K.H.}, \bibinfo{author}{Kim, M.U.}, \bibinfo{year}{1986}.
\newblock \bibinfo{title}{The onset of natural convection in a fluid layer suddenly heated from below}.
\newblock \bibinfo{journal}{Int. J. Heat Mass Transfer} \bibinfo{volume}{29}, \bibinfo{pages}{193--201}.
\bibitem[{Kim et~al.(1999)Kim, Choi and Choi}]{Kim99ijhmt}
\bibinfo{author}{Kim, M.C.}, \bibinfo{author}{Choi, K.H.}, \bibinfo{author}{Choi, C.K.}, \bibinfo{year}{1999}.
\newblock \bibinfo{title}{The onset of thermal convection in an initially, stably stratified fluid layer}.
\newblock \bibinfo{journal}{Int. J. Heat Mass Transfer} \bibinfo{volume}{42}, \bibinfo{pages}{4253--4258}.
\bibitem[{Kim et~al.(2002)Kim, Park and Choi}]{Kim02tcfd}
\bibinfo{author}{Kim, M.C.}, \bibinfo{author}{Park, H.K.}, \bibinfo{author}{Choi, C.K.}, \bibinfo{year}{2002}.
\newblock \bibinfo{title}{Stability of an initially stably stratified fluid subjected to a step change in temperature}.
\newblock \bibinfo{journal}{Theoret. Comp. Fluid Dyn.} \bibinfo{volume}{16}, \bibinfo{pages}{49--57}.
\bibitem[{Kim et~al.(1996)Kim, Yoon and Choi}]{Kim96kjce}
\bibinfo{author}{Kim, M.C.}, \bibinfo{author}{Yoon, D.Y.}, \bibinfo{author}{Choi, C.K.}, \bibinfo{year}{1996}.
\newblock \bibinfo{title}{Buoyancy-driven convection in a horizontal fluid layer under uniform volumetric heat sources}.
\newblock \bibinfo{journal}{Korean J. Chem. Eng.} \bibinfo{volume}{13}, \bibinfo{pages}{165--171}.
\bibitem[{Mahler et~al.(1968)Mahler, Schechter and Wissler}]{Mahler68pf}
\bibinfo{author}{Mahler, E.G.}, \bibinfo{author}{Schechter, R.S.}, \bibinfo{author}{Wissler, E.H.}, \bibinfo{year}{1968}.
\newblock \bibinfo{title}{Stability of a fluid layer with time-dependent density gradients}.
\newblock \bibinfo{journal}{Phys. Fluids} \bibinfo{volume}{11}, \bibinfo{pages}{1901--1912}.
\bibitem[{Mizushima(1995)}]{Mizushima95jpsj}
\bibinfo{author}{Mizushima, J.}, \bibinfo{year}{1995}.
\newblock \bibinfo{title}{Onset of convection in a finite two-dimensional box}.
\newblock \bibinfo{journal}{J. Phys. Soc. Jpn.} \bibinfo{volume}{64}, \bibinfo{pages}{2420--2432}.
\bibitem[{Plevan and Quinn(1966)}]{Plevan66aj}
\bibinfo{author}{Plevan, R.E.}, \bibinfo{author}{Quinn, J.A.}, \bibinfo{year}{1966}.
\newblock \bibinfo{title}{The effect of monomolecular films on the rate of gas absorption into a quiescent liquid}.
\newblock \bibinfo{journal}{AIChE J.} \bibinfo{volume}{12}, \bibinfo{pages}{894--902}.
\bibitem[{Rayleigh(1916)}]{Rayleigh16pm}
\bibinfo{author}{Rayleigh, L.}, \bibinfo{year}{1916}.
\newblock \bibinfo{title}{On convection currents in a horizontal layer of fluid when the higher temperature in on the under side}.
\newblock \bibinfo{journal}{Philos. Mag.} \bibinfo{volume}{32}, \bibinfo{pages}{529}.
\bibitem[{Spangenberg and Rowland(1961)}]{Spangenberg61pf}
\bibinfo{author}{Spangenberg, W.G.}, \bibinfo{author}{Rowland, W.R.}, \bibinfo{year}{1961}.
\newblock \bibinfo{title}{Convective circulation in water induced by evaporative convection}.
\newblock \bibinfo{journal}{Phys. Fluids} \bibinfo{volume}{4}, \bibinfo{pages}{743--750}.
\bibitem[{Sparrow et~al.(1964)Sparrow, Goldstein and Jonsson}]{Sparrow64jfm}
\bibinfo{author}{Sparrow, E.M.}, \bibinfo{author}{Goldstein, R.J.}, \bibinfo{author}{Jonsson, V.K.}, \bibinfo{year}{1964}.
\newblock \bibinfo{title}{Thermal instability in a horizontal fluid layer: Effect of boundary conditions and non-linear temperature profile}.
\newblock \bibinfo{journal}{J. Fluid Mech.} \bibinfo{volume}{18}, \bibinfo{pages}{513--528}.
\bibitem[{Stull(1988)}]{Stull88book}
\bibinfo{author}{Stull, R.B.}, \bibinfo{year}{1988}.
\newblock \bibinfo{title}{An Introduction to Boundary Layer Meteorology}.
\newblock \bibinfo{edition}{5th} ed., \bibinfo{publisher}{Kluwer Academic Publishers}.
\bibitem[{Tan and Thorpe(1992)}]{Tan92ces}
\bibinfo{author}{Tan, K.K.}, \bibinfo{author}{Thorpe, R.B.}, \bibinfo{year}{1992}.
\newblock \bibinfo{title}{Gas diffusion into viscous and non-{N}ewtonian liquids}.
\newblock \bibinfo{journal}{Chem. Eng. Sci.} \bibinfo{volume}{47}, \bibinfo{pages}{3565--3572}.
\bibitem[{Ueda et~al.(1984)Ueda, Komori, Miyasaki and Ozoe}]{Ueda84pf}
\bibinfo{author}{Ueda, H.}, \bibinfo{author}{Komori, S.}, \bibinfo{author}{Miyasaki, S.}, \bibinfo{author}{Ozoe, H.}, \bibinfo{year}{1984}.
\newblock \bibinfo{title}{Time-dependent thermal convection in a stably stratified fluid layer heated from below}.
\newblock \bibinfo{journal}{Phys. Fluids} \bibinfo{volume}{27}, \bibinfo{pages}{2617--2623}.
\bibitem[{Yang and Choi(2002)}]{Yang02pf}
\bibinfo{author}{Yang, D.J.}, \bibinfo{author}{Choi, C.K.}, \bibinfo{year}{2002}.
\newblock \bibinfo{title}{Onset of thermal convection in a horizontal fluid layer heated from below with time-dependent heat flux, the}.
\newblock \bibinfo{journal}{Phys. Fluids} \bibinfo{volume}{14}, \bibinfo{pages}{930--937}.

\end{thebibliography}

\end{document}